\documentclass[journal,twoside,final]{IEEEtran}
\usepackage{graphicx}
\usepackage{amsmath}
\usepackage{latexsym}
\usepackage{graphicx}
\usepackage{bm}
\usepackage{amssymb}
\usepackage{array}
\usepackage{setspace}
\usepackage{fancyhdr}
\usepackage{amsmath, amssymb, epsfig}
\usepackage{float}
\usepackage{subcaption}
\usepackage{balance}
\usepackage{color}

\usepackage{csquotes}
\MakeOuterQuote{"}

\makeatletter
\newcommand{\vast}{\bBigg@{3.2}}
\newcommand{\Vast}{\bBigg@{4.2}}
\makeatother

\newtheorem{theorem}{Theorem}

\def\proof{\noindent\hspace{2em}{\itshape Proof: }}
\def\endproof{\hspace*{\fill}~$\blacksquare$\par\endtrivlist\unskip}

\begin{document}
%
\title{\linespread{1} Impact of Interference on the Performance of RIS-Assisted Source DF Relaying Networks}

\author{Doaa Salim\authorrefmark{1}, Monjed~H.~Samuh\authorrefmark{1},~and~Anas~M.~Salhab\authorrefmark{2},~\IEEEmembership{Senior~Member,~IEEE}\\
\thanks{\authorrefmark{1} Doaa Salim and Monjed H. Samuh are with the Department of Applied Mathematics \& Physics,
Palestine Polytechnic University, Hebron, Palestine (e-mails: doasleem-1995@hotmail.com, monjedsamuh@ppu.edu).}
\thanks{\authorrefmark{2} Anas M. Salhab is with the Department
of Electrical Engineering, King Fahd University of Petroleum \&
Minerals, Dhahran 31261, Saudi Arabia (e-mail: salhab@kfupm.edu.sa).}}


\maketitle{}

\begin{abstract}
This letter investigates the impact of co-channel interference (CCI) on the performance of a decode-and-forward (DF) relaying network with a reconfigurable intelligent surface (RIS)-assisted source. We consider one source, multiple DF relays, and one destination with CCI at both the relays and destination. We derive closed-form accurate approximations for the system outage probability assuming Rayleigh fading channels. In addition, we study the system performance at the high signal-to-noise ratio (SNR) regime, where closed-form expressions are derived for the asymptotic outage probability, diversity order, and the coding gain. The results show that the number of reflecting elements $N$ at the source has a small effect on the coding gain of the system and not on the diversity order. Furthermore, findings illustrate that the number of relays $K$ is affecting the diversity order and is more impactful on the performance than $N$. Finally, results show that utilizing RIS at the source node mitigates the interference effect at the relay nodes.  

\end{abstract}
\textbf{\small \emph{Index Terms\textemdash}Reconfigurable intelligent surface, decode-and-forward relay, Rayleigh fading, co-channel interference.}


%
\IEEEpeerreviewmaketitle


\section{Introduction}
Recently and due to their capability to enhance the system performance, the reconfigurable intelligent surfaces (RISs) have attracted a noticeable attention as a promising candidate for future wireless communication networks. 
An RIS is an artificial surface, made of electromagnetic
material, that is capable of customizing the propagation of
the radio waves impinging upon it \cite{Renzo1}, \cite{Alouini1}. 
It has been proposed as a new low-cost and less complicated solution to realize wireless
communication with high spectrum and energy efficiencies. 
  
An overview of the basic characteristics of the large intelligent
surface/antenna technology and its potential applications has been provided in \cite{Liang}. Furthermore, a detailed 
overview on the state-of-the-art solutions, fundamental differences of RIS with other 
technologies, and the most important open research issues in this area of research has been 
provided by authors recently in \cite{Basar1}.

In \cite{Wu}, it has been shown that RIS has better performance than
conventional massive multiple-input multiple-output (MIMO)
systems as well as better performance than multi-antenna
amplify-and-forward (AF) relaying networks with smaller
number of antennas, while reducing the system complexity and
cost. Recently, Yang \textit{et al.} studied in \cite{Yang1} the performance 
of RIS-assisted mixed indoor visible light communication (VLC)/radio frequency (RF) system.
They derived closed-form expressions for the outage probability and bit error rate (BER) for AF 
and decode-and-forward (DF) relaying schemes. A study that compares the performance of relay-assisted and RIS-assisted
wireless networks from coverage, probability of signal-to-noise ratio (SNR) gain, and delay outage rate has been
provided in \cite{Yang2}.

The outage probability and BER performance of a dual-hop mixed free space optical (FSO)-RF relay 
network with RIS has been studied in \cite{Yang3}. In \cite{Yang4}, the authors utilized RIS to improve 
the quality of a source signal that is sent to destination through an unmanned aerial vehicle (UAV). The average BER of a RIS-assisted network with space-shift keying has been recently studied in \cite{Canbilen}. In \cite{Yang5}, analytical expression has been derived for the secrecy outage probability of RIS-assisted network in the presence of direct link and eavesdropper.

It is important to mention here that most of the previous works on RIS-assisted networks performed their analysis based on the central limit theorem (CLT), which makes them applicable only for large number of reflecting
elements. Following other approaches and to cover the case of low number of reflecting elements, the authors in \cite{Yang6} and \cite{Boulogeorgos} have derived accurate approximations for the channel distributions and performance metrics of RIS-assisted networks assuming Raleigh fading channels. Recently, some works on RIS-assisted networks over Nakagami-$m$ fading channels started to appear in literature \cite{Ferreira}, \cite{Anas1}. Most recently, limited number of papers have considered the interference phenomenon in a RIS context \cite{Li}, \cite{Hou}. All these works on interference considered the scenario, where an RIS is used as a relay node.

As can be seen, the topic of RIS-assisted networks is still open for research and attracting many researchers to investigate it from various aspects. Motivated by this, we consider in this letter a different scenario, where a source is utilizing RIS to enhance the transmitted signal quality in a DF relaying network with co-channel interference (CCI). We derive accurate closed-form approximations for the system outage probability assuming Rayleigh fading channels for both intended users and interferers. The derived results are valid for arbitrary number of reflecting elements $N$. In addition, in order to get more insights at the system performance, we derive closed-form expression for the asymptotic outage probability at high SNR values, where the system diversity order and coding gain are provided and analyzed. To the best of authors’ knowledge, the derived expressions are new in an RIS-aided wireless system context. Furthermore, the achieved key findings on the impact of RIS on the system performance are being reported for the first time in this letter.


\section{System and Channel Model}\label{SM}
Consider a dual-hop relay system with RIS-aided source of $N$ reflecting elements, $K$ DF relay nodes, one destination, and arbitrary number of interferers at both the relays and destination with opportunistic relay selection scheme. The entire communication takes place in two phases. In the first phase, the source {\sf S} transmits its signal to $K$ relays. In the second phase, only the best relay among all other relays who succeeded in decoding the source signal in the first phase is selected to forward it to destination {\sf D}. We assume that the signal at the $k^{\mathrm{th}}$ relay is corrupted by interfering signals from
$I_{k}$ co-channel interferers $\{x_{i}\}_{i=1}^{I_{k}}$.

The received signal at the $k^{\mathrm{th}}$ relay can be expressed as
\begin{equation}\label{Eq.1}
y_{{\sf r}_{k}}= \sum_{i = 1}^{N} h_{{\sf s},k,i}x_{0}+\sum_{i_{k}=1}^{I_{k}}h_{ i_{k},k}^{I}x_{i_{k},k}^{I}+n_{{\sf
s},k},
\end{equation}
where $h_{{\sf s},k,i}$ is the channel coefficient between the $i^{\mathrm{th}}$ reflecting element at $\sf S$ and the $k^{\mathrm{th}}$ relay, $x_{0}$ is the transmitted symbol with $\mathbb{E}\{|{x_{0}}|^{2}\}=P_{0}$, $h_{i_{k},k}^{I}$ is the channel coefficient between the $i_{k}^{\mathrm{th}}$ interferer and $k^{\mathrm{th}}$ relay, $x_{i_{k},k}^{I}$ is the transmitted symbol from the $i_{k}^{\mathrm{th}}$ interferer with
$\mathbb{E}\{|{x_{i_{k},k}^{I}}|^{2}\}=P_{i_{k},k}^{I}$, $n_{{\sf s},k}\thicksim\mathcal{CN}(0, N_{0})$ is an additive white Gaussian noise (AWGN) of zero mean and power $N_{0}$, and $\mathbb{E}\{\cdot\}$ denotes the expectation operation. Let us define ${h_{k,\sf d}}$, and ${h_{i_{d},\sf d}^{I}}$ as the channel coefficients between the $k^{\mathrm{th}}$ relay and {\sf
D}, and the $i_{d}^{\mathrm{th}}$ interferer and {\sf D}, respectively. The channel coefficients between {\sf S} and the $k^{\mathrm{th}}$ relay ${h_{{\sf s},k,i}}, k=1...K$ are assumed to be Rayleigh distributed with mean $\frac{\sqrt{\pi}}{2}$ and variance $\frac{(4-\pi)}{4}$. That is, their mean powers $\mathbb{E}\{|h_{{\sf s},k,i}|^{2}=1\}$. In addition, all the other channel coefficients are assumed to follow Rayleigh distribution. That is, the channel powers denoted by $|{h_{k,\sf d}}|^2$, $|{h_{i_{k},k}^{I}}|^2$, and $|{h_{i_{d},\sf d}^{I}}|^2$ are
exponential distributed random variables (RVs) with parameters $\sigma_{k,\sf d}^{2}$, $\sigma_{I,i_{k},k}^{2}$, and
$\sigma_{I,i_{d},\sf d}^{2}$, respectively. Using (\ref{Eq.1}),
the signal-to-interference-plus-noise ratio (SINR) at the $k^{\mathrm{th}}$ relay can be written as
\begin{equation}
\gamma_{{\sf s},k}=\frac{P_{0}}{N_{0}}{\left(\sum_{i=1}^{N}|h_{{\sf
s},k,i}|\right)^{2}}\Big/\left(\sum_{i_{k}=1}^{I_{k}}\frac{P_{i_{k},k}^{I}}{N_{0}}|{h_{
i_{k},k}^{I}}|^2+1\right).
\end{equation}

Let $B_{L}$ denote a decoding set defined by the set of relays who
successfully decoded the source message at the first phase. It is
defined as
\begin{align}\label{Eq.3}
B_{L}&\triangleq\left\{k\in \mathcal{S}_{r}:\gamma_{{\sf s},k}\geq
2^{2R}-1\right\},
\end{align}
where $\mathcal{S}_{r}$ is the set of all relays and $R$ denotes a
fixed spectral efficiency threshold.

In the second phase and after decoding the received signal, only the best relay in $B_{L}$ forwards the
re-encoded signal to the destination. The best
relay is the relay with the maximum
$\gamma_{l,\sf d}$, where $\gamma_{l,\sf d}$ is the SINR at the destination resulting from the $l^{\mathrm{th}}$ relay being the
relay, which forwarded the source information. It can be written as
\begin{equation}
\gamma_{l,\sf d}=\frac{P_{l}}{N_{0}}|{h_{l,\sf
d}}|^2\Big/\left(\sum_{i_{d}=1}^{I_{d}}\frac{P_{i_{d},\sf
d}^{I}}{N_{0}}\big|{h_{ i_{d},\sf d}^{I}}\big|^2+1\right),
\end{equation} where
$P_{l}$, $P_{i_{d},\sf d}^{I}$, and $N_{0}$ are the transmit power
of the $l^{\mathrm{th}}$ active relay, the transmit power of the $i_{d}^{\mathrm{th}}$ interferer, and the AWGN power at the
destination, respectively, and $I_{d}$ is the number of
interferers at the destination node. Since the denominator is
common to the SINRs from all relays belonging to $B_{L}$, the
best relay is the relay with the
maximum $\left\{\frac{P_{l}}{N_{0}}|{h_{l,\sf
d}}|^2\right\}$.

The end-to-end (e2e) SINR
at {\sf D} can be written as
\begin{equation}\label{dsinr}
\gamma_{\sf
d}=\frac{P_{b}}{N_{0}}|{h_{b,\sf
d}}|^2\bigg/\left(\sum_{i_{d}=1}^{I_{d}}\frac{P_{i_{d},\sf
d}^{I}}{N_{0}}\big|{h_{ i_{d},\sf d}^{I}}\big|^2+1\right),
\end{equation}
where the subscript $b$ is used in \eqref{dsinr} to denote the best selected relay.
\section{Outage Performance Analysis}\label{PA}
In this section, we derive a closed-from approximation for the system
outage probability. Before going into details, we find it is
appropriate to first present some preliminary studies, and hence,
the new results of the considered system can be revealed.
\subsection{Preliminary Study}
The probability of the decoding set defined in (\ref{Eq.3}) can be
written as
\begin{align}\label{Eq.8}
\mathrm{P_{r}}\left[B_{L}\right]=\prod_{l\in
B_{L}}\mathrm{P_{r}}\left[\gamma_{{\sf s},l}\geq
u\right]\prod_{m\notin B_{L}}\mathrm{P_{r}}\left[\gamma_{{\sf
s},m}< u\right],
\end{align}
where $u=\left(2^{2R}-1\right)$ is the outage SNR threshold.
The outage probability of the system can be achieved by averaging
over all the possible decoding sets as follows \cite{JKim}
\begin{align}\label{Eq.9}
P_{\sf out}&\triangleq\mathrm{P_{r}}\left[\frac{1}{2}\
\mathrm{log}_{2}\left(1+\gamma_{\sf d}\right)<R\right]\nonumber\\
&=\sum_{L=0}^{K}\sum_{B_{L}}\mathrm{P_{r}}\left[\gamma_{\sf
d}<u|B_{L}\right]\mathrm{P_{r}}\left[B_{L}\right],
\end{align}
where the internal summation is taken over all of
${{K}\choose{L}}$ possible subsets of size $L$ from the set with
$K$ relays. In order to evaluate (\ref{Eq.9}), we need first to
derive $\mathrm{P_{r}}\left[\gamma_{\sf d}<u|B_{L}\right]$ and
$\mathrm{P_{r}}\left[B_{L}\right]$, which are presented in the
following section.

Throughout the analysis below, it assumed that $\rho=
P_{0}/N_{0}=P_{l}/N_{0}$ and $\rho_{I}=
P_{i_{k},k}^{I}/N_{0}=P_{i_{d},\sf d}^{I}/N_{0}$. The terms $\rho|h_{l,\sf d}|^{2}$,
$\rho_{I}|h_{i_{k},k}^{I}|^{2}$, and
$\rho_{I}|h_{i_{d},\sf d}^{I}|^{2}$ are exponential distributed
with parameters $\lambda_{l,\sf d}=1/\rho\sigma_{l,\sf d}^{2}$, $\lambda_{i_{k},k}^{I}=1/\rho_{I}\sigma_{I,i_{k},k}^{2}$, and
$\lambda_{i_{d},\sf d}^{I}=1/\rho_{I}\sigma_{I,i_{d},\sf d}^{2}$. As the channels between {\sf S} and all relays are assumed to have the same mean power, which equals 1, their average SNRs will be equal to $\rho=
P_{0}/N_{0}$ and their parameters will be equal to $\lambda_{{\sf s},k}=1/\rho, k=1,...,K$.

\subsection{Outage Probability}
The approximate outage probability of the considered system is summarized in the
following key result.

\begin{theorem}\label{theorem:1}
The outage probability of RIS-assisted source DF relaying network with independent identically distributed (i.i.d.) second hop cumulative distribution functions (CDFs) ($\lambda_{1,{\sf d}}=\lambda_{2,{\sf d}}=...=\lambda_{{\sf r,d}}$) and i.i.d. interferers' powers at {\sf D} ($\lambda^{I}_{1,{\sf d}}=\lambda^{I}_{2,{\sf d}}=...=\lambda^{I}_{I_{d},{\sf d}}=\lambda^{I}_{{\sf d}}$) and at relays ($\lambda^{I}_{1,k}=\lambda^{I}_{2,k}=...=\lambda^{I}_{I_{k},k}=\lambda^{I}_{k}, k=1,...,K$) can be obtained in a closed-form expression by using \eqref{Eq.9}, after evaluating the
terms $\mathrm{P_{r}}\left[\gamma_{\sf d}<u|B_{L}\right]$ and
$\mathrm{P_{r}}\left[\gamma_{{\sf s},k}<u\right]$ as follows
\end{theorem}
\begin{align}\label{Eq.27az}
&\mathrm{P_{r}}\left[\gamma_{\sf
d}<u|B_{L}\right]= - \frac{\left(\lambda_{\sf
 d}^{I} \right)^{I_{d}}}{\left(I_{d} - 1 \right)!}e^{\lambda_{\sf r,d}^{I}} \left(-1 \right)^{I_{d}}\sum_{g=0}^{I_{d} - 1}\ \begin{pmatrix}
I_{d} - 1 \\ g
\end{pmatrix} (-1)^{g} \nonumber\\&\times\sum_{i=0}^{L} \begin{pmatrix}
L \\ i \end{pmatrix} (-1)^{i}  \left( \lambda^{I}_{\sf d}+ \lambda_{\sf r,d} u i \right)^{-g - 1} \Gamma\left(g+1, \lambda^{I}_{\sf d}+ \lambda_{\sf r,d} u i\right),
\end{align}
\begin{align}\label{Eq.27bz}
&\mathrm{P_{r}}\left[\gamma_{{\sf
s},k}<u\right]\approx \frac{\left(\lambda_{k}^{I} \right)^{I_{k}}}{\left(I_{k} - 1 \right)!}e^{\lambda_{k}^{I}} \left(-1 \right)^{I_{k}}\sum_{g=0}^{I_{k} - 1}\ \begin{pmatrix}
I_{k} - 1 \\ g
\end{pmatrix} (-1)^{g} \nonumber\\
 &\times \vast(- (\lambda^{I}_{k})^{-g-1} \Gamma\left(g+1,\lambda^{I}_{k}\right)+ \sum_{i = 0}^{N - 1} \frac{u^{i}  (\lambda_{{\sf s},k})^{i}}{B^i i!} \nonumber\\
 &\times\left(\lambda^{I}_{k} + \frac{u  \lambda_{{\sf s},k}}{B}\right)^{-g-i-1}\Gamma\left(g+i+1,\lambda^{I}_{k} + \frac{u \lambda_{{\sf s},k}}{B }\right)\vast),
\end{align}
where $\Gamma(.,.)$ denotes the upper incomplete gamma function \cite[Eq. (8.352.2)]{Grad.}.

\proof  See Appendix \ref{appendix:1}.\endproof

%
%
\section{Asymptotic Outage Behavior}\label{AOB}
In this section, we evaluate the system performance at high SNR
values in which the outage probability can be expressed as $P_{\sf
out}\approx\left(G_{c}\rho\right)^{-G_{d}}$, where $G_{c}$ is the coding gain of the system and $G_{d}$ is the diversity order \cite{Alouinib}.

\begin{theorem}\label{Theorem:2}
The asymptotic outage probability for RIS-assisted source DF relaying network with i.i.d. second hop CDFs and i.i.d. interferers at both the relays and destination can be obtained in a closed-form expression by using \eqref{Eq.9}, after evaluating the
terms $\mathrm{P_{r}}\left[\gamma_{\sf d}<u|B_{L}\right]$ and
$\mathrm{P_{r}}\left[\gamma_{{\sf s},k}<u\right]$ as follows
\begin{align}\label{Eq.28a}
&\mathrm{P_{r}}\left[\gamma_{\sf
d}<u|B_{L}\right]\approx - \frac{\left(\lambda_{\sf
 d}^{I} \right)^{I_{d}}}{\left(I_{d} - 1 \right)!}e^{\lambda_{\sf d}^{I}} \left(-1 \right)^{I_{d}}\sum_{g=0}^{I_{d} - 1}\ \begin{pmatrix}
I_{d} - 1 \\ g
\end{pmatrix} (-1)^{g} \nonumber\\&\times (\lambda_{\sf r,d})^{L} u^{L} (\lambda_{\sf d}^{I})^{-g-L-1} \Gamma\left(g+L+1, \lambda^{I}_{\sf d}\right),
\end{align}
\begin{align}\label{Eq.28b}
&\mathrm{P_{r}}\left[\gamma_{{\sf
s},k}<u\right]\approx - \frac{\left(\lambda_{k}^{I} \right)^{I_{k}}e^{\lambda_{k}^{I}} \left(-1 \right)^{I_{k}}}{\left(I_{k} - 1 \right)!B^{N}N!}\sum_{g=0}^{I_{k} - 1}\ \begin{pmatrix}
I_{k} - 1 \\ g
\end{pmatrix} (-1)^{g} \nonumber\\&\times (\lambda_{{\sf s},k})^{N} u^{N} (\lambda_{k}^{I})^{-g-N-1} \Gamma\left(g+N+1, \lambda^{I}_{k}\right).
\end{align}
\end{theorem}
\proof See Appendix \ref{appendix:3}.\endproof 

After evaluating the asymptotic outage probability and studying it carefully, it has been noticed that we have two extreme cases for the system performance as follows. 
\begin{itemize}
\item Case 1: $N=1$\\
In this case, the outage probability is dominated by the first hop of the system, namely the term $\mathrm{P_{r}}\left[\gamma_{{\sf s},k}<u\right]$. Therefore, the asymptotic outage probability can be expressed as
\begin{align}\label{Eq.28bss}
&P_{\sf out}^{\infty}\approx \vast(\vast[- \frac{\left(\lambda_{\sf r}^{I} \right)^{I_{r}}e^{\lambda_{\sf r}^{I}} \left(-1 \right)^{I_{r}}}{\left(I_{r} - 1 \right)!B^{N}N!}\sum_{g=0}^{I_{r} - 1}\ \begin{pmatrix}
I_{r} - 1 \\ g
\end{pmatrix}\nonumber\\
&\times (-1)^{g} u  \frac{\Gamma\left(g+2, \lambda^{I}_{\sf r}\right)}{(\lambda_{\sf r}^{I})^{g+2}}\vast]^{-1}\rho\vast)^{-K}.
\end{align}
It is clear from \eqref{Eq.28bss} that the diversity order of the system for this case is $G_{d}=K$ and the coding gain $G_{c}$ is a function of several parameters, including $N$, $I_{\sf r}$, $\lambda_{\sf r}^{I}$, and $u$.\\

\item Case 2: $N\geq 2$\\
In this case, the outage probability is dominated by the second hop of the system, namely the term $\mathrm{P_{r}}\left[\gamma_{\sf
d}<u|B_{L}\right]$. Therefore, the asymptotic outage probability can be expressed as
\begin{align}\label{Eq.28ass}
&P_{\sf out}^{\infty}\approx \vast(\vast[- \frac{\left(\lambda_{\sf d}^{I} \right)^{I_{d}}}{\left(I_{d} - 1 \right)!}e^{\lambda_{\sf d}^{I}} \left(-1 \right)^{I_{d}}\sum_{g=0}^{I_{d} - 1}\ \begin{pmatrix}
I_{d} - 1 \\ g
\end{pmatrix}\nonumber\\
&\times (-1)^{g} u^{L}  \frac{\Gamma\left(g+L+1, \lambda^{I}_{\sf d}\right)}{(\lambda_{\sf d}^{I})^{g+L+1}}\vast]^{-1/L}\rho\vast)^{-L}.
\end{align}
The term $\mathrm{P_{r}}\left[\gamma_{\sf
d}<u|B_{L}\right]$ is dominant when the number of active relays is equal to number of all available relays, that is $L=K$. This means that the diversity order of this system for this case is also equal to $G_{d}=K$, whereas it is clear from \eqref{Eq.28ass} that the coding gain $G_{c}$ is a function of several parameters, including $I_{d}$, $\lambda_{\sf d}^{I}$, and $u$. It is important to notice here that there still exists an effect for $N$ in this case, which comes from the first hop terms. This effect is on the coding gain of the system, but it is not easy to quantify and reflect it in \eqref{Eq.28ass} as it is coming from large number of terms, especially when $K$ is large. 
\end{itemize}

It is worthwhile to mention here again that there exists a minor effect in Case 2, which comes from increasing $N$. This effect is more noticeable when $N$ is quite smaller than $K$. It is important also to mention here that changing $N$ from 1 to 2 always has an impact on the coding gain of the system regardless of the value of $K$. For the cases where $N\geq 2$, the gain achieved in $G_{c}$ becomes less as $N$ becomes closer to $K$ till it vanishes. In addition, all the cases where $N\geq K-1$ always converge to the same performance at the very high values of SNR. 
\section{Numerical Results}\label{NR}
For simplicity, we assume here that the parameters $\sigma_{k,\sf d}^{2}$, $\sigma_{I,i_{k},k}^{2}$, and
$\sigma_{I,i_{d},\sf d}^{2}$ are all equal to 1. In addition, we call the outage SNR threshold as $\gamma_{\sf out}=u$.

\begin{figure}[htb!]
\centering
\includegraphics[scale=0.34]{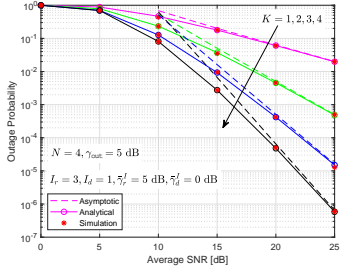}
\caption{$P_{\sf out}$ vs SNR for different values of $K$.}\label{Pout_SNR_K_New}
\end{figure}

Fig. \ref{Pout_SNR_K_New} validates the achieved analytical results. It
can be seen that the analytical results as well as the asymptotic
curves perfectly fit with the simulation ones. We can also notice
that the outage probability decreases as $K$ increases.
Furthermore, it is obvious that the diversity order linearly
increases with the number of relays $K$ although we only use
one relay.

\begin{figure}[htb!]
\centering
\includegraphics[scale=0.34]{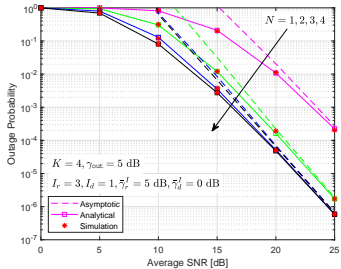}
\caption{$P_{\sf out}$ vs SNR for different values of $N$.}\label{Pout_SNR_N_New}
\end{figure}

Fig. \ref{Pout_SNR_N_New} illustrates the impact of number of reflecting elements $N$ at the source on the system performance. Clearly, a good matching is happening between the analytical and asymptotic results with the simulation ones in this figure. In addition, the two cases studied in the asymptotic analysis section are clear in this figure. We have the case of $N=1$ and the case of $N\geq 2$. The diversity order for both cases is $G_{d}=K$, whereas the coding gain $G_{c}$ is a function of the interference parameters at the relay nodes and the outage threshold for the case where $N=1$, and is a function of the interference parameters at the destination node and the outage threshold for the case where $N\geq 2$ with a minor effect of $N$. It is also clear that as $N$ becomes close to $K$, the gain achieved in the coding gain, and hence, the system performance becomes smaller till it vanishes when $N$ becomes $\geq K-1$.
\restylefloat{figure}

\begin{figure}[htb!]
\centering
\includegraphics[scale=0.34]{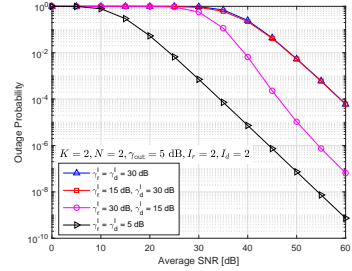}
\caption{$P_{\sf out}$ vs SNR for different values of $\bar{\gamma}_{\sf r}^{I}$ and $\bar{\gamma}_{\sf d}^{I}$.}\label{Pout_SNR_gIr_gId}
\end{figure}

Fig. \ref{Pout_SNR_gIr_gId} studies the impact of interference on the system performance. It is obvious from this figure that the interference at the relays has a negligible effect on the system performance, where clearly the case of $\bar{\gamma}_{\sf r}^{I}=15\ \mathrm{dB}$ and $\bar{\gamma}_{\sf d}^{I}=30\ \mathrm{dB}$ gives almost the same performance as the case of $\bar{\gamma}_{\sf r}^{I}=\bar{\gamma}_{\sf d}^{I}=30\ \mathrm{dB}$ with a very small enhancement at the low range of average SNR. In contrast, it clear that the interference at the destination is noticeably impacting the system performance where the case of $\bar{\gamma}_{\sf r}^{I}=30\ \mathrm{dB}$ and $\bar{\gamma}_{\sf d}^{I}=15\ \mathrm{dB}$ is outperforming the case of $\bar{\gamma}_{\sf r}^{I}=\bar{\gamma}_{\sf d}^{I}=30\ \mathrm{dB}$. This enhancement in the performance comes through the coding gain $G_{c}$. Compared to results in \cite{Anas}, we can see here that the presence of RIS at the source reverses the effect of interference at the relays and destination on the system performance, where the interference at the destination becomes more impactful. In addition, we can see that utilizing RIS mitigates the impact of interference at the relays. Finally, the best performance is achieved at the case of $\bar{\gamma}_{\sf r}^{I}=\bar{\gamma}_{\sf d}^{I}=5\ \mathrm{dB}$, where decreasing $\bar{\gamma}_{\sf d}^{I}$ from 15 dB to 5 dB is again enhancing the coding gain of the system.

\begin{figure}[htb!]
\centering
\includegraphics[scale=0.34]{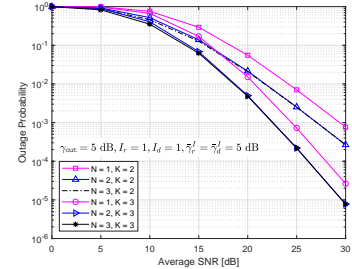}
\caption{$P_{\sf out}$ vs SNR for different values of $N$ and $K$.}\label{Pout_SNR_N_K_New}
\end{figure}

Fig. \ref{Pout_SNR_N_K_New} illustrates the impact of $K$ and $N$ on the system performance. We can see from this figure that $K$ is affecting the diversity order of the system, where changing $K$ from 2 to 3 increased $G_{d}$ from 2 to 3. On the other hand, it is clear that changing $N$ affects the coding gain $G_{c}$ of the system and not the diversity order. We can also see that regardless of the value of $K$, increasing $N$ from 1 to 2 always enhances $G_{c}$ with a clear amount. On the other hand, we can see that the gain achieved in the system performance when $N$ keeps increasing becomes very minor as $N$ becomes close to $K$. In addition, at very high values of SNR, all the cases where $N\geq K-1$ converge to the same performance.   
\restylefloat{figure}

\section{Conclusion}\label{C}
This letter investigated the impact of CCI on the performance of a DF relaying network with a RIS-assisted source. Closed-form approximate expression was derived for the system outage probability assuming Rayleigh fading channels. In addition, the system performance was studied at the high SNR regime, where closed-form expressions were derived for the asymptotic outage probability, diversity order, and the coding gain. The results showed that $N$ has a small effect on $G_{c}$ and not on $G_{d}$. Furthermore, findings illustrated that $K$ is affecting $G_{d}$ and is more impactful on the performance than $N$. Finally, results showed that utilizing RIS at the source node mitigates the interference effect at the relay nodes.

\appendices
\section{Proof of Theorem \ref{theorem:1}}\label{appendix:1}
In this Appendix, we evaluate the first term in (\ref{Eq.9})
$\mathrm{P_{r}}\left[\gamma_{\sf d}<u|B_{L}\right]$. First, the e2e SINR
can be written as $\gamma_{\sf d} = Y/Z$. The CDF of
$\gamma_{\sf d}$ given a decoding set $B_{L}$ is given by
$\mathrm{P_{r}}\left[\gamma_{\sf
d}<u|B_{L}\right]=\int_{1}^{\infty}f_{Z}(z)\int_{0}^{uz}f_{Y}(y)dydz = \int_{1}^{\infty}f_{Z}(z)F_{Y}(uz)dz$.
As mentioned before, with Rayleigh distributed channel coefficients, the SNR CDFs of the second hops and interferers at {\sf D} will be following exponential distribution as $F_{\gamma_{\sf r,d}}(\gamma)=1-e^{-\lambda_{\sf r,d}\gamma}$ and $F_{\gamma_{\sf d}^{I}}(\gamma)=1-e^{-\lambda_{\sf d}^{I}\gamma}$, respectively. Now, the CDF of $Y$, which is the CDF of the best selected relay can be expressed with the help of the binomial rule as
\begin{align*}
F_{Y}(uz)&= (F_{\gamma_{\sf r,d}}(uz))^L=( 1 - e^{-\lambda_{\sf r,d}u z})^L\\
&= \sum_{i = 0}^{L} \begin{pmatrix}
L \\ i 
\end{pmatrix}  (-1)^i \exp\left(-\lambda_{\sf r,d}u z i\right).
\end{align*}
The RV $Z$ is equal to $Z=X+1$, where $X$ is the interference at the destination, which has the following PDF \cite{Anas}
\begin{align}\label{Eq.13w}
f_{X}(x)= - \frac{\left(\lambda_{\sf
 d}^{I} \right)^{I_{d}}}{\left(I_{d} - 1 \right)!}x^{I_{d}-1}e^{-\lambda_{\sf d}^{I}x}.
\end{align}
Using transformation of RVs and then the binomial rule, the PDF of $Z$ can be obtained as
\begin{align}\label{Eq.13}
f_{Z}(z)= - \frac{\left(\lambda_{\sf
 d}^{I} \right)^{I_{d}}}{\left(I_{d} - 1 \right)!}e^{\lambda_{\sf d}^{I}} \left(-1 \right)^{I_{d}}\sum_{g=0}^{I_{d} - 1}\ \begin{pmatrix}
I_{d} - 1 \\ g
\end{pmatrix} (-1)^{g} z^{g} e^{-\lambda_{\sf d}^{I}z}.
\end{align}
Now, using the integral $\int_{1}^{\infty}f_{Z}(z)F_{Y}(uz)dz$ and with the help of \cite[Eq. (8.351.2)]{Grad.}, the first term in \eqref{Eq.9} $\mathrm{P_{r}}\left[\gamma_{\sf d}<u|B_{L}\right]$ can be obtained as in \eqref{Eq.27az}.

To obtain the second term in (\ref{Eq.9}) $\mathrm{P_{r}}\left[\gamma_{{\sf s},k}<u\right]$, the first hop SINR can be written as $Y'/Z'$, where $Y'$ is the first hop SNR with RIS-aided transmitter and $Z'=X'+1$, where $X'$ is the interference at the relays. Again, the CDFs of the interferers at the relays follow exponential distribution as $F_{\gamma_{k}^{I}}(\gamma)=1-e^{-\lambda_{k}^{I}\gamma}, k=1...K$. The CDF of first hop SNR is given by \cite{Yang6}
\begin{align}
F_{Y'}(uz) = 1- e^{-\frac{\lambda_{{\sf s},k} u z}{B}} \sum_{i = 0}^{N - 1} \frac{(\lambda_{{\sf s},k} u z)^i}{B^i i!},\end{align}
where $B = 1 + (N - 1)\Gamma^{2}\left(\frac{3}{2} \right)$.\\
Again, upon using the integral $\int_{1}^{\infty}f_{Z'}(z)F_{Y'}(uz)dz$ and with the help of \cite[Eq. (8.351.2)]{Grad.}, the second term in \eqref{Eq.9} $\mathrm{P_{r}}\left[\gamma_{{\sf s},k}<u\right]$ can be obtained as in \eqref{Eq.27bz}.

\section{Proof of Theorem \ref{Theorem:2}}\label{appendix:3}
To find the asymptotic outage probability, we first need to
obtain $\mathrm{P_{r}}\left[\gamma_{\sf d}<u|B_{L}\right]$. As
$\rho\rightarrow\infty$ and with constant values of $\rho_{I}$,
$I_{k}$, and $I_{d}$, and using the Taylor series expansion, the CDF of the exponential
distribution becomes $\lambda u$, and hence, the CDF of the second hop channels with opportunistic relaying can be approximated as
$F_{Y}(uz)\approx (\lambda_{{\sf r,d}} u z)^L$. With no change in the PDF of $Z$ and upon following the same procedure as in Appendix \ref{appendix:1},
the term $\mathrm{P_{r}}\left[\gamma_{\sf d}<u|B_{L}\right]$ in
(\ref{Eq.9}) can be evaluated at high SNR values as in \eqref{Eq.28a}.

Now, the second term in \eqref{Eq.9}
$\mathrm{P_{r}}\left[B_{L}\right]$ can be obtained after
evaluating the CDF of $\gamma_{{\sf s},k}$, which can be
approximated at high SNR values as
\begin{align}\label{C.1}
F_{\gamma_{{\sf s},k}}(u z)\approx
\frac{(u z)^N}{(B/ \lambda_{{\sf s},k})^{N} N!}.
\end{align}
Again, with no change in the PDF of $Z'$ and upon following the same procedure as in Appendix \ref{appendix:1},
the term $\mathrm{P_{r}}\left[B_{L}\right]$ in
(\ref{Eq.9}) can be evaluated at high SNR values as in \eqref{Eq.28b}.


\balance




\end{document}